\documentclass[a4paper,useAMS,usenatbib]{mn2e}

\usepackage{ifthen,psfig,graphicx}
\usepackage{epstopdf,aas_macros}
\usepackage{soul,color,amsmath}
\usepackage{rotating,longtable,caption}
\usepackage{dblfloatfix}
\usepackage{placeins}
\usepackage{multirow}
\usepackage{caption}
\usepackage{subfigure}
\def\mathnew{\mathsurround=0pt}
\def\simov#1#2{\lower 2.5pt\vbox{\baselineskip0pt \lineskip-.5pt
\ialign{$\mathnew#1\hfil##\hfil$\crcr#2\crcr\sim\crcr}}}

\newcommand{\MeV}{Me\kern-0.11em V}
\newcommand{\keV}{ke\kern-0.11em V}

\newcommand{\Lsun}{\ensuremath{{\rm L}_{\odot}}}

\newcommand{\Msun}{\ensuremath{{\rm M}_{\odot}}}

\makeatletter
\newcommand\ion[2]{\mbox{#1$\;${\small\expandafter\@slowromancap\romannumeral #2@\relax}}}
\newcommand\iont[2]{{#1$\;${\small\expandafter\@slowromancap\romannumeral #2@\relax}}}
\makeatother
\makeatletter
\newcommand{\raisemath}[1]{\mathpalette{\raisem@th{#1}}}
\newcommand{\raisem@th}[3]{\raisebox{#1}{$#2#3$}}

\def\mathnew{\mathsurround=0pt}
\def\simov#1#2{\lower 2.5pt\vbox{\baselineskip0pt \lineskip-.5pt
\ialign{$\mathnew#1\hfil##\hfil$\crcr#2\crcr\sim\crcr}}}
\def\lesssim{\mathrel{\mathpalette\simov <}}
\def\gtrsim{\mathrel{\mathpalette\simov >}}

\makeatother

\title[Follow-up observations of 2XMMi~J184725.1$-$631724]{Multiwavelength Follow-up Observations of the Tidal Disruption Event Candidate 2XMMi~J184725.1$-$631724}
\author[D. Lin et al.]{Dacheng Lin$^{1}$\thanks{E-mail:
    dacheng.lin@unh.edu}, Jay Strader$^{2}$, Eleazar
  R. Carrasco$^{3}$, Olivier Godet$^{4,5}$, Dirk Grupe$^{6}$,
  \newauthor   Natalie A. Webb$^{4,5}$,  Didier Barret$^{4,5}$, Jimmy A. Irwin$^{7,8}$\smallskip\\
  $^{1}$Space Science Center, University of New Hampshire, Durham, NH
  03824, USA\\
  $^{2}$Center for Data Intensive and Time Domain Astronomy, Department of Physics and Astronomy, Michigan State University,\\
  567 Wilson Road, East Lansing, MI 48824, USA\\
$^{3}$Gemini Observatory/AURA, Southern Operations Center, Casilla
603, La Serena, Chile\\
$^{4}$CNRS, IRAP, 9 avenue du Colonel Roche, BP 44346, F-31028
Toulouse Cedex 4, France\\
$^{5}$Universit\'{e} de Toulouse, UPS-OMP, IRAP, Toulouse, France\\
$^{6}$Space Science Center, Morehead State University, 235 Martindale Drive,
Morehead, KY 40351, USA\\
$^{7}$Department of Physics and Astronomy, University of Alabama, Box 870324, Tuscaloosa, AL 35487, USA\\
$^{8}$Department of Physics and Astronomy, Seoul National University, Seoul 08826, Korea\\
}
\begin{document}

\date{In original form 2017 June 10}

\pagerange{\pageref{firstpage}--\pageref{lastpage}} \pubyear{2017}

\maketitle

\label{firstpage}

\begin{abstract}

The ultrasoft X-ray flare 2XMMi~J184725.1$-$631724 was serendipitously
detected in two \emph{XMM-Newton} observations in 2006 and 2007, with
a peak luminosity of $6\times10^{43}$~erg~s$^{-1}$. It was suggested
to be a tidal disruption event (TDE) because its position is
consistent with the center of an inactive galaxy. It is the only known
X-ray TDE candidate whose X-ray spectra showed evidence of a weak
steep powerlaw component besides a dominant supersoft thermal disk. We
have carried out multiwavelength follow-up observations of the
event. Multiple X-ray monitorings show that the X-ray luminosity has
decayed significantly after 2011. Especially, in our deep
\emph{Chandra} observation in 2013, we detected a very faint
counterpart that supports the nuclear origin of 2XMMi~J184725.1$-$631724
but had an X-ray flux a factor of $\sim$1000 lower than in the
peak of the event. Compared with follow-up UV observations, we found
that there might be some enhanced UV emission associated with the TDE
in the first \emph{XMM-Newton} observation. We also obtained a
high-quality UV-optical spectrum with the SOAR and put a very tight
constraint on the persistent nuclear activity, with a persistent X-ray
luminosity expected to be lower than the peak of the flare by a factor
of $>$2700. Therefore, our multiwavelength follow-up observations
strongly support the TDE explanation of the event.

\end{abstract}


\section{Introduction}
\label{sec:intro}
Most supermassive black holes (SMBHs) at the center of galaxies are
quiescent, but they can sometimes reveal themselves by tidally
disrupting and subsequently accreting stars that wander too close to
them \citep{re1988,re1990,ko2012,ko2015}. About 70 such tidal
disruption events (TDEs) have been found thus far, with about 30 with
X-ray detections \citep[e.g.,][]{koba1999,gechre2012,sarees2012,mauler2013,mikami2015}\footnote{https://tde.space/}. While three hard TDEs
discovered by \emph{Swift} had peak X-ray luminosity $L_\mathrm{X}$
above $10^{48}$~erg~s$^{-1}$ and hard X-ray spectra
\citep{blgime2011,bukegh2011,cekrho2012,brlest2015}, most X-ray TDEs
have peak $L_\mathrm{X}$ $\lesssim$10$^{44}$~erg~s$^{-1}$ and pure
thermal X-ray spectra of characteristic temperature $\lesssim$0.1
keV. Some TDEs, such as 2XMMi~J184725.1$-$631724 \citep[J1847
  hereafter,][Lin11 hereafter]{licagr2011} and XMMSL1~J074008.2-853927
\citep{sareko2017}, seemed to have more complicate X-ray spectra, with
an extra harder component besides the super-soft thermal component.

J1847 is among the four X-ray TDE candidates that we have discovered
\citep[Lin11,][]{liirgo2013,limair2015,ligoho2017,liguko2017} in our
project of classification of bright X-ray sources from the
\emph{XMM-Newton} catalog \citep{liweba2012}. It was serendipitously
detected in two \emph{XMM-Newton} observations (hereafter X1 and X2,
Table~\ref{tbl:obslog}) in the direction of the center of the galaxy
IC 4765-f01-1504 \citep{camein2006} at a redshift of 0.0353 ($D_L=149$
Mpc, assuming a flat universe with $H_0$=73 km s$^{-1}$ Mpc$^{-1}$ and
$\Omega_{\rm M}$=0.27, Lin11). These two \emph{XMM-Newton}
observations were separated by 211~d, with the X-ray flux in X2
higher than in X1 by nearly one order of magnitude. The X-ray spectra
in both observations are very soft and can be described by a model
dominated by a cool (temperature $\lesssim$0.1 keV) thermal disk
component plus a weak steep (photon index
$\Gamma_\mathrm{PL}$$\sim$3.5) power-law (PL), similar to the thermal
state of stellar-mass black hole (BH) X-ray binaries
\citep{remc2006}. The large amplitude (by a factor $>$64) of the flare
is supported by the non-detection of the source in a \emph{ROSAT}
observation in 1992. The decay of the X-ray flare had been hinted by
the non-detection of the source in a follow-up {\it Swift} observation
$\sim$4 yr after X2, which implied a flux decaying factor of
$\gtrsim$12. The host galaxy showed no clear sign of persistent
nuclear activity, based on our Gemini spectrum taken in 2011, which
showed no emission lines but typical stellar absorption features. The
probable transient nature, extremely soft X-ray spectra, and an
otherwise quiescent host prompted Lin11 to conclude that it is a
TDE. The X-ray spectra of the TDE candidate XMMSL1~J074008.2-853927
are different from those of J1847 in that X-ray spectra of
XMMSL1~J074008.2-853927 are more similar to those typically seen in
Seyfert galaxies, with the PL component being hard
($\Gamma_\mathrm{PL}$$\sim$2) and strong \citep[relative to the soft
  component,][]{sareko2017}.

We have obtained more follow-up observations of J1847 from optical to
X-rays, including one new \textit{XMM-Newton} observation (X3
  hereafter, Tables~\ref{tbl:obslog} and \ref{tbl:obsloguv}), one
  \textit{Chandra} observation (C1 hereafter), five new \emph{Swift}
  observations (S2--S6 hereafter), and one SOAR optical spectrum.
Here we report the results of these new observations, which we find to
further support the TDE nature of the event. In
Section~\ref{sec:reduction}, we describe the observations and data
analysis. In Section~\ref{sec:res}, we present the results. Our
discussion on the source nature and conclusions are given in
Section~\ref{sec:discussion}.

\begin{table*}
\setlength\tabcolsep{3pt} 
\centering \caption{X-ray observations of J1847.} \label{tbl:obslog}
\begin{tabular}{lllllllccc}
\hline
Observatory & Observation\,ID & Start date & Instrument & Exp$^{a}$  & $r_\mathrm{src}^{b}$ & Count Rate$^{c}$ & $F_{\rm abs}$$^{d}$ & $F_{\rm unabs}$$^{d}$ & $L_\mathrm{bol}$$^{d}$ \\
&&  & & (ks)  & (arcsec) & ($10^{-3}$ cts s$^{-1}$) &\multicolumn{2} {c} {($10^{-13}$~erg~s$^{-1}$ cm$^{-2}$)}  & ($10^{43}$~erg~s$^{-1}$)\\
\hline
\textit{XMM-Newton} & 0405550401(X1)  & 2006-09-06 & pn &19.5 &15 &$102\pm2$& $2.2\pm0.1$ & $18\pm3$ & $1.81\pm0.57$\\
&& & MOS1 &27.6 &$15$  &$21.5\pm0.9$\\
&& & MOS2 &27.6 &$15$ &$17.9\pm0.8$\\
&0405380501(X2) & 2007-04-16 & pn & 20.5 &35  &$865\pm6$& $19.2\pm0.3$ & $102\pm9$ & $6.40\pm0.72$ \\
&& & MOS1 & 32.2 &35  & $159\pm2$\\
&& & MOS2 & 31.9 &35  & $122\pm2$\\
&0694610101(X3)$^{e,f}$  & 2013-03-03 & pn&19.1& 20 &  $<$6.7 & $<$0.25 &$<$0.31 & $<$0.98 \\
&& & MOS1 &25.6 & 20 &$<$1.5\\
&& & MOS2 &25.5 & 20  &$<$3.1\\
\hline
\textit{Swift}& 00031930001(S1) & 2011-02-23 &XRT& 5.0& 20& $<$3.3 &$<$1.5 &$<$19 &  $<$6.1\\
&00031930002(S2)$^{e}$ &2013-04-05 &XRT& 3.1 & 20& $<$2.1 &$<$1.2 & $<$14 & $<$7.6\\
&00031930003(S3)$^{e}$ &2013-04-07 &XRT&2.8 & 20 \\
&00031930004(S4)$^{e}$ & 2013-08-29  &XRT&1.3 & 20 & $<$3.4 &$<$1.5 &$<$18 & $<$9.5 \\
&00031930005(S5)$^{e}$ &2013-09-08 &XRT &1.8 & 20\\
&00031930006(S6)$^{e}$ & 2015-03-15  &XRT& 1.4 & 20 & $<$5.6 & $<$2.1 &$<$23 & $<$18.8\\
\hline
\textit{Chandra}&15637(C1)$^{e}$ & 2013-03-29 & ACIS-S& 14.9 &1.1 & $0.32\pm0.15$ &$0.020^{+0.020}_{-0.012}$ & $0.24^{+0.24}_{-0.15}$ & $0.13^{+0.13}_{-0.08}$\\
\hline
\textit{ROSAT}&800256(R1)& 1992-10-11& PSPC &10.7 &$40$ &$<$2.3 &$<$0.63 &$<$0.87 &$<$0.037\\
\hline
\end{tabular}
\begin{list}{}{}
\item[$^{a}$]\textit{XMM-Newton} observations are all subject to background flares, the exposure times given are the clean exposure times after excluding periods of strong background flares.
  \item[$^{b}$]The radius of the source extraction region.
\item[$^{c}$]The count rates correspond to different energy bands for different observatories (\textit{XMM-Newton} in 0.2--10 keV, \textit{Swift} in 0.3--10 keV, \textit{Chandra} in 0.3--8 keV, and \textit{ROSAT} in 0.1--2.4 keV), with $1\sigma$ errors or $3\sigma$ upper limits (calculated with the \textsc{ciao} task \texttt{aprates}).
\item[$^{d}$]The 0.2--10 keV absorbed ($F_{\rm abs}$) and unabsorbed ($F_{\rm unabs}$) fluxes and the unabsorbed bolometric luminosity $L_\mathrm{bol}$ (integrated over 0.001--100 keV) for X1 and X2 are taken from Lin11, based on the WABS*SIMPL(DISKBB) model. For the \textit{ROSAT} observation R1, we assumed an absorbed PL of $\Gamma_\mathrm{PL}=2.0$ and $N_\mathrm{H}=8.5\times10^{20}$ cm$^{-2}$ (the bolometric luminosity were integrated over 0.2--100 keV for this model). For all the other observations, we assumed the WABS*SIMPL(DISKBB) fit to X1, with the disk temperature adjusted based on the expected decrease in the bolometric luminosity (see the text for more details). Either 90\% errors or $3\sigma$ upper limits are given.
\item[$^{e}$]New follow-up observations.
  \item[$^{f}$]For X3, the observed count rates, fluxes and the bolometric luminosity are treated as upper limits, given that in this observation J1847 was strongly contaminated by Source 2.
\end{list}
\end{table*}

\begin{table*}
\centering \caption{The UV Photometry} \label{tbl:obsloguv}
\begin{tabular}{lccccccc}
\hline
 Observation\,ID & Start date & Filter  &$\lambda_\mathrm{eff}$ &Exposure &Magnitude\\
& & & (\AA)& (ks) & (AB mag)  \\
\hline
\multicolumn{2} {l} {\textit{XMM-Newton}/OM:}\\
0405550401(X1) & 2006-09-06 & $UVW1$  &2910& 1.8 & $19.78\pm0.09$\\
           &            & $UVM2$  &2310& 2.7 & $19.90\pm0.11$\\
0694610101(X3)$^{a}$ & 2013-03-03 & $UVW1$  &2910& 9.8 & $20.34\pm0.06$\\
&            & $UVM2$  &2310& 9.6 & $20.46\pm0.08$\\
           &            & $UVW2$  &2120&14.3 & $20.62\pm0.13$\\
\hline
\multicolumn{2} {l} {\textit{Swift}/UVOT:}\\
00031930001(S1) & 2011-02-23 & $UVW1$ & 2589 & 4.9 & $20.08\pm0.05$\\
00031930002(S2)$^{a}$ & 2013-04-05 & $UVW1$ & 2589 & 3.1 & $20.19\pm0.07$ \\
00031930003(S3)$^{a}$ & 2013-04-07 & $UVW2$ & 2031 & 2.8 & $20.42\pm0.06$\\
00031930004(S4)$^{a}$ & 2013-08-29 & $UVW2$ & 2031 & 1.3 &$20.27\pm0.08$ \\
00031930005(S5)$^{a}$ & 2013-09-08 & $UVW2$ & 2031 & 1.8 &$20.36\pm0.07$ \\
00031930006(S6)$^{a}$ & 2015-03-15 & $UVW2$ & 2031 & 1.4 &$20.60\pm0.09$ \\
\hline
\end{tabular}
\begin{list}{}{}
\item[$^{a}$]New follow-up observations.
\end{list}
\end{table*}

\section{DATA ANALYSIS}
\label{sec:reduction}
\subsection{The \emph{XMM-Newton} and \emph{Chandra} Observations}

The \textit{XMM-Newton} observation X3 was taken on 2013 March 3
(Table~\ref{tbl:obslog}) with a duration of 44 ks using the three
European Photon Imaging Cameras \citep[i.e., pn, MOS1, and
MOS2][]{jalual2001,stbrde2001,tuabar2001} in Full Frame
mode. Strong background flares, which were seen in all
  cameras, were excluded following the \textsc{sas} thread for the filtering
  against high
  backgrounds\footnote{http:\/\/xmm.esac.esa.int\/sas\/current\/documentation\/threads\/}.
After excluding the strong background intervals, we were left with
only good exposures of 19.1, 25.6 and 25.5 ks for pn, MOS1, and
MOS2, respectively (Table~\ref{tbl:obslog}). There was some weak hard
emission near the position of J1847. However, this hard source seems
to be significantly offset from J1847. The offset is 4.8~arcsec and is
larger than the $3\sigma$ positional uncertainty (3.4~arcsec) of the
hard source, based on the 3XMM-DR6 catalog \citep{rowewa2016}. In
order to check whether this hard source is J1847, we requested the
\emph{Chandra} observation C1 (15 ks, Table~\ref{tbl:obslog}) in the
director's discretionary time program. The observation was made on
2013 March 29 (26~d after X3). It used the AXAF CCD Imaging
Spectrometer \citep[ACIS; ][]{bapiba1998}, with the aimpoint and our
target at the back-illuminated chip S3.  We reprocessed the data to
apply the latest calibration (CALDB 4.7.3) using the script
\texttt{chandra\_repro} in the {\it Chandra} Interactive Analysis of
Observations (\textsc{ciao}, version 4.9) package. No high background
  flares were seen in C1, and we used the data of the whole
  observation.

The \textit{Chandra} image clearly resolved the emission near J1847 in
X3 into two sources, one at the position of J1847, thus probably the
same source, and the other only 5.9~arcsec away (Source 2
hereafter). This offset was obtained based on our performance of the
source detection on the 0.3--7 keV image binned at single sky pixel
resolution with the \textsc{ciao} \texttt{wavdetect} wavelet-based source
detection algorithm \citep{frkaro2002}. We extracted the spectra and
the corresponding response files for both sources using the \textsc{ciao} task
\texttt{specextract}. We used a circular source region enclosing 90\%
of the point spread function (PSF) at 2.0 keV (radius 1.1~arcsec for
both sources) and a circular background region of radius 20~arcsec
near the sources.

Although J1847 cannot be resolved from Source 2 in X3, the observation
can still provide enough statistics to constrain the properties of
these sources, thanks to the long exposure and the large effective
area of the observation. Therefore, we analyzed the observation in the
same way as for X1 and X2 (Lin11), except using \textsc{sas} 16.0.0 and the
calibration files of 2017 February. We extracted the source and
background spectra for the blended emission of J1847 and Source 2 in
X3. We used a source region of radius 20~arcsec (the same for all
cameras) centered at the detection position in the 3XMM-DR6 catalog. A
large background region (radius 100~arcsec for MOS1 and MOS2 and
radius 60~arcsec for pn) was used.

\subsection{\emph{Swift} Observations}

We also have six follow-up observations of J1847 with \emph{Swift},
including the one in 2011 that had been analyzed in Lin11
(Table~\ref{tbl:obslog}). We analyzed the six observations with
\textsc{ftools} 6.20 and the calibration files of 2017 February. In
all observations, the X-ray telescope \citep[XRT;][]{buhino2005} was
operated in Photon Counting mode, and we reprocessed the data with the
task \texttt{xrtpipeline} (version 0.13.3) to update the
calibration. J1847 was not detected in the XRT in these
observations. In order to obtain the constraint of the X-ray flux of
the source in these observations, we extracted the source and
background spectra using circular source and background regions of
radii 20 and 100~arcsec, respectively. The $3\sigma$
upper limit on the count rate was estimated using the \textsc{ciao}
task \texttt{aprates}, which adopts the Bayesian approach. The
UV-Optical Telescope \citep[UVOT;][]{rokema2005} in these observations
used the UV filters $UVW1$ and $UVW2$ (Table~\ref{tbl:obsloguv}), with
a goal to monitor the UV emission. We used the task
\texttt{uvotsource} with radii of 5 and 20~arcsec for the
circular source and background regions, respectively, and the most
recent UVOT calibration as described in \citet{pobrpa2008} and
\citet{becuho2010,brlaho2011} to obtain the UV photometry.

\subsection{The SOAR Spectroscopic Observation}
We obtained a new optical spectrum of the host galaxy of J1847 using
the Goodman High-Throughput Spectrograph \citep{clcran2004} on the
SOAR 4.1-m telescope on 2013 August 26. Seven separate 30 min
exposures were obtained, interspersed with arc lamps and flats, for a
total on-source exposure time of 3.5 hr.  For all data, we used the
400 l mm$^{-1}$ grating, centered at about 5000 \AA, for an
approximate wavelength coverage of 3000 to 7000 \AA. We used a
1.07-arcsec slit, giving a resolution of 6.8 \AA. The
  individual exposures were reduced and optimally extracted, using
  apertures ranging from 2.07 to 2.31~arcsec, depending on the seeing in
  each exposure. Then the spectra from individual exposures were
  coadded to yield the final spectrum.

We used Penalized Pixel Fitting (\textsc{ppxf}) software \citep{caem2004}
to fit the spectrum with multi-component models comprised of
single-population synthetic spectra from \cite{vasafa2010}, spanning a
grid of 48 ages between 0.06 to 14 Gyr and 7 metallicities
[M/H]=\{$-2.32$, $-1.71$, $-1.31$, $-0.71$, $-0.40$, $0.00$,
$+0.22$\}. The \textsc{ppxf} software implements the penalized
pixel-fitting method to extract the stellar kinematics or stellar
population from spectra of galaxies, with a maximum
penalized likelihood approach.  A multiplicative polynomial of 10
degrees was incorporated in the fit to account for possible calibration
uncertainties. The spectrum was corrected for the Galactic dust
reddening of $\mathrm{E}(B-V)_\mathrm{G}=0.10$ mag \citep{scfida1998}
before the fit.

\subsection{Astrometry Correction}
We need to obtain the astrometrically corrected positions of J1847 in
various X-ray images to check whether the source is consistent with
the galactic origin. The X1 and X2 sources were astrometrically
corrected by aligning the 3XMM-DR6 sources with the 2MASS astrometry
using 4 and 6 X-ray-IR matches, respectively (J1847 was not in these
match lists used for astrometry correction). We used the astrometric
correction method in \citet{licawe2016}, by searching for the
translation and rotation of the X-ray frame that minimize the total
$\chi^2$ of the matches ($\chi$ is the ratio of the X-ray-IR
separation to the total positional error). The uncertainties of the
translation and rotation and thus the systematic positional errors of
the X-ray sources associated with the astrometry correction procedure
were estimated using 200 simulations. The final source positional
uncertainties include the statistical component, the systematic error
associated with the astrometry correction procedure, and the
systematic error of 0.37~arcsec ($1\sigma$) of the 3XMM-DR6 catalog
\citep{rowewa2016}, added in quadrature. We note that the
contamination of Source 2 has negligible effects on the position of
J1847 that we derived from X1 and X2. This is because Source 2 in
these observations would have count rates lower than those of J1847 by
factors of 14 and 130, respectively, if it had similar fluxes as
observed in C1.

The positions of the sources in C1 were taken from the output of the source
detection tool \texttt{wavdetect}. The positional uncertainties included
the statistical component, calculated using Equation 12 in
\citet{kikiwi2007}, and the absolution astrometric uncertainty of
\emph{Chandra} obtained in \citet{robu2011}. No large systematic
astrometric offset was found for C1 when we applied the astrometric
correction method of \citet{licawe2016} to compare it with the 2MASS
sources.

In order to check the galactic origin of J1847, we used an
  archival Very Large Telescope (VLT) FORS1 $R$-band image, which
  resolves the candidate host galaxy of J1847 well. It was taken on
  2006 April 16. We aligned the astrometry of the image with that of
  the 2MASS. The VLT image has a relatively small size of $6.8\arcmin
  \times 6.8\arcmin$, which is the reason why we did not use it to
  carry out astrometry correction for X-ray sources.

\section{Results}
\label{sec:res}
\subsection{Deep X-ray Follow-up}
\label{sec:xraypos}

\begin{figure}
\centering
\subfigure{
  \includegraphics[width=3.2in]{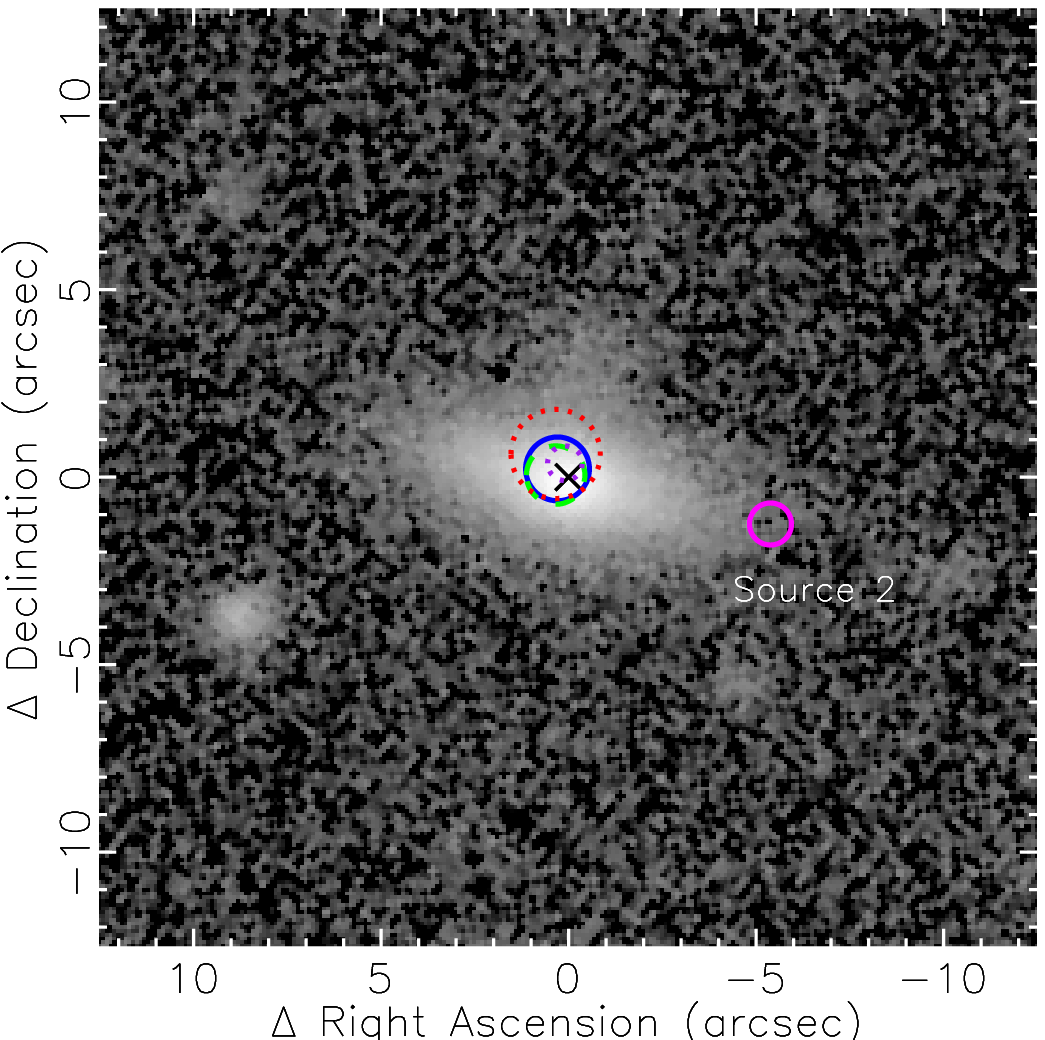}
}
\subfigure{\includegraphics[width=3.2in]{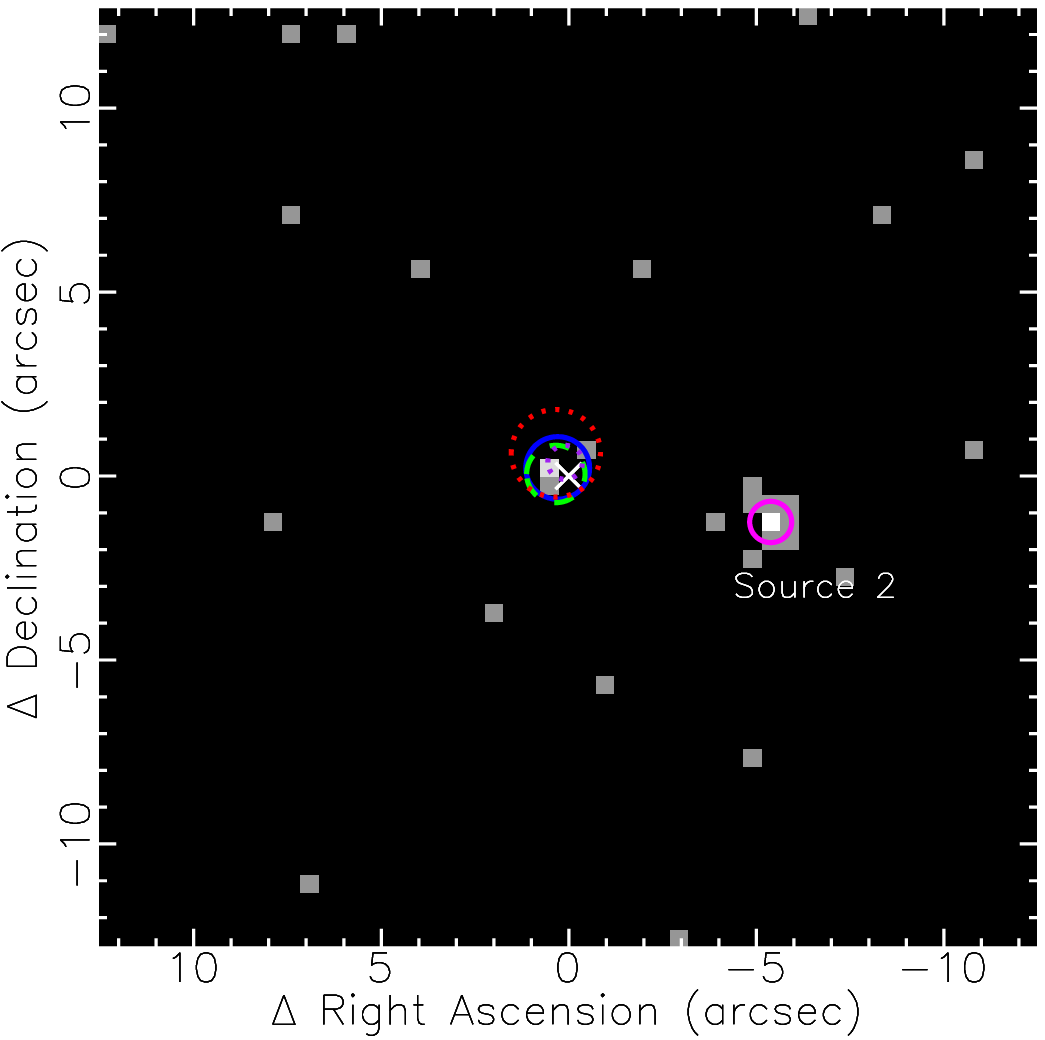}
  }
\caption{The VLT FORS1 $R$-band image (upper panel) and the
  \emph{Chandra} X-ray image (C1, lower panel) around the field of
  J1847. The origins of the images are at the center (black cross in
  the upper panel and white cross in the lower panel,
  R.A.=18$^\mathrm{h}$47$^\mathrm{m}$25$^\mathrm{s}$.12,
  decl.=$-63^\circ$17$\arcmin$25$\farcs$3) of its host galaxy. The
  blue solid circle (radius 0.85~arcsec, i.e., 0.6 kpc,
  R.A.=18$^\mathrm{h}$47$^\mathrm{m}$25$^\mathrm{s}$.16,
  decl.=$-63^\circ$17$\arcmin$25$\farcs$1) represents the 95\%
  positional uncertainty of J1847 from the \emph{Chandra} observation
  C1. The green dashed circle (radius 0.78~arcsec, i.e., 0.5 kpc,
  R.A.=18$^\mathrm{h}$47$^\mathrm{m}$25$^\mathrm{s}$.17,
  decl.=$-63^\circ$17$\arcmin$25$\farcs$2) is from the combination of
  \emph{XMM-Newton} X-ray observations X1 and X2. The red dotted
  circle (radius 1.19~arcsec, i.e. 0.8 kpc,
  R.A.=18$^\mathrm{h}$47$^\mathrm{m}$25$^\mathrm{s}$.17,
  decl.=$-63^\circ$17$\arcmin$24$\farcs$7) is from the UV observation in X1. The purple
  dotted circle (radius 0.46~arcsec, i.e., 0.3 kpc,
  R.A.=18$^\mathrm{h}$47$^\mathrm{m}$25$^\mathrm{s}$.13,
  decl.=$-63^\circ$17$\arcmin$24$\farcs$9) is from the stacked $UVW2$
  image of the \emph{Swift} observations. The magenta solid circle
  (radius 0.56~arcsec,
  R.A.=18$^\mathrm{h}$47$^\mathrm{m}$24$^\mathrm{s}$.31,
  decl.=$-63^\circ$17$\arcmin$26$\farcs$6) marks the 95\% positional
  uncertainty of Source 2, a nearby source detected in C1. \label{fig:optimg}}
\end{figure}

\begin{figure}
\centering
\includegraphics[width=3.4in]{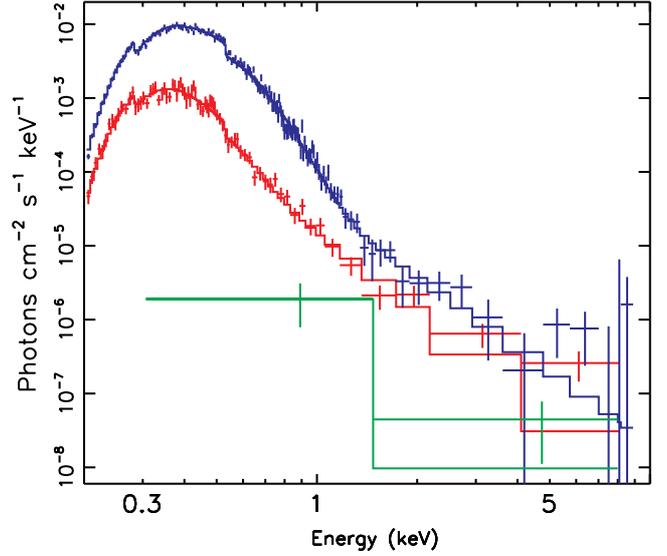}
\caption{The X-ray spectra of J1847 from X2, X1, and C1 (from
    the top to the bottom). The X1 and X2 spectra were fitted with the
    WABS*SIMPL(DISKBB) model (from Lin11). For clarity, only pn
    spectra are shown. The model for C1 is the same as that of X2,
    except that the disk temperature was set to 35 eV (see the
    text). \label{fig:spfits}}
\end{figure}

In the \emph{Chandra} observation C1, we collected 5 counts and 17
counts for J1847 and Source 2, respectively, within \mbox{0.3--8~keV},
with only 0.2 counts expected to be from the background in both
cases. The 95\% positional uncertainties of both sources from this
\emph{Chandra} observation are marked in Figure~\ref{fig:optimg},
which shows the VLT FORS1 $R$-band image (upper panel) and the
\emph{Chandra} image (lower panel) around the field of J1847. The
\emph{Chandra} position of J1847 is consistent with the galaxy center
within the 95\% uncertainty. It is also consistent with the average
position (weighted by errors) from X1 and X2
(Figure~\ref{fig:optimg}).

The very few counts of J1847 in the \emph{Chandra} observation
  do not allow for a meaningful fit even with a simple model like an
  absorbed PL. The absorbed X-ray flux $F_\mathrm{abs}$ within 0.2--10
  keV is fairly independent of the spectral shape. Assuming an absorbed
  PL of $N_\mathrm{H}=8.5\times10^{20}$ cm$^{-2}$ (Lin11) and a large
  range of $\Gamma_\mathrm{PL}$ (2.0--5.0), we obtained
  $F_\mathrm{abs}=$ (1.5--2.0)$\times10^{-15}$~erg~s$^{-1}$ cm$^{-2}$,
  a factor of $\sim$1000 lower than the peak flux (in X2,
  Table~\ref{tbl:obslog}). The corresponding unabsorbed 0.2--10 flux
  had a much larger range ((2--20)$\times10^{-15}$~erg~s$^{-1}$ cm$^{-2}$). 

The bolometric luminosity $L_\mathrm{bol}$ is more relevant
  for the test of the TDE explanation for J1847, but it is very
  sensitive to the spectral model adopted. In Lin11, the X1 and X2
  spectra were well fitted with the WABS*SIMPL(DISKBB) model, as
  plotted in Figure~\ref{fig:spfits}. The absorption model WABS, the
  standard thermal disk model DISKBB and the empirical convolution
  Comptonization model SIMPL, in which a fraction of the input seed
  photons are converted into a PL \citep{stnamc2009}, can be found in
  the X-ray spectral fitting package \textsc{xspec} \citep{ar1996}. Based on
  this model, the main spectral change between X1 and X2 was the inner
  disk temperature $kT_\mathrm{disk}$, as often observed in the
  thermal state of the accreting stellar-mass BH X-ray
  binaries \citep{remc2006,dogiku2007}. Therefore, we could construct
  a reasonable WABS*SIMPL(DISKBB) model for the C1 spectrum of J1847,
  as long as its $L_\mathrm{bol}$ and thus $kT_\mathrm{disk}$ can be
  inferred in some way. In the standard TDE theory, $L_\mathrm{bol}$
  is expected to decay with the time $t$ after the stellar disruption
  as $t^{-5/3}$ \citep{re1988,ph1989}. The solid line in
  Figure~\ref{fig:ltlc} plots such a decay trend with the disruption
  time assumed to be one month before X1, which is reasonable, given
  that the rising time to the peak is expected to be $\sim$1--2 months
  \citep{ul1999,gura2015} for a small BH ($\sim$$10^6$ \Msun,
  Lin11). Based on this model, we expect
  $L_\mathrm{bol}=1.5\times10^{42}$~erg~s$^{-1}$ at the time of
  C1. This luminosity can be matched if we adjust the
  WABS*SIMPL(DISKBB) fit to X1 to have $kT_\mathrm{disk}=35$ eV. The
  model constructed this way is fully consistent with the C1 spectrum,
  as shown in Figure~\ref{fig:spfits}. The model predicts 3.3 counts,
  while we detected 4.8 net counts in C1. 

Source 2 has enough counts for simple spectral fits. We binned the
0.3--8 keV spectrum to have at least 1 count per bin and fitted it
with the $C$ statistic \citep{ca1979,waleke1979} in \textsc{xspec} with an
absorbed PL model. We obtained $\Gamma_\mathrm{PL}=0.0^{+1.2}_{-0.7}$
(the error bars of the parameters from the spectral fits are at the
90\% confidence level throughout the paper),
$N_\mathrm{H}=0.0^{+1.7}\times10^{22}$ cm$^{-2}$, and a 0.3--8 keV
absorbed flux of $f_\mathrm{0.3-0.8\
  keV}=2.8^{+1.7}_{-1.3}\times10^{-14}$ erg cm$^{-2}$ s$^{-1}$,
indicating a faint X-ray source of a very hard spectrum.

In order to check whether there is large spectral and/or flux
variability between X3 and C1, we also fitted the combined spectrum of
J1847 and Source 2 from C1 with an absorbed PL and obtained
$N_\mathrm{H}=0.0^{+0.8}\times10^{22}$ cm$^{-2}$,
$\Gamma_\mathrm{PL}=0.3\pm0.6$, and
$f_\mathrm{0.3-0.8\ keV}=3.2_{-1.3}^{+2.0}\times10^{-14}$ erg
cm$^{-2}$ s$^{-1}$.  These values are broadly consistent with those
inferred in the fit to the combined 0.2--10 keV spectrum of J1847 and
Source 2 from X3 ($N_\mathrm{H}=0.1^{+0.1}\times10^{22}$~cm$^{-2}$,
$\Gamma_\mathrm{PL}=1.1\pm0.3$, and
$f_\mathrm{0.3-0.8\ keV}=3.5\pm0.8\times10^{-14}$~erg~cm$^{-2}$~s$^{-1}$).
These parameters are broadly consistent with each other. Therefore,
there should be no large flux and spectral change of both sources
between X3 and C1. However, given the relatively large
uncertainties of these parameters, we cannot rule out small variations
like a factor of two in the total X-ray flux.

\subsection{Long-term X-ray Evolution}
\label{sec:ltxraycurve}

\begin{figure}
  \centering
  \subfigure{
    \includegraphics[width=3.4in]{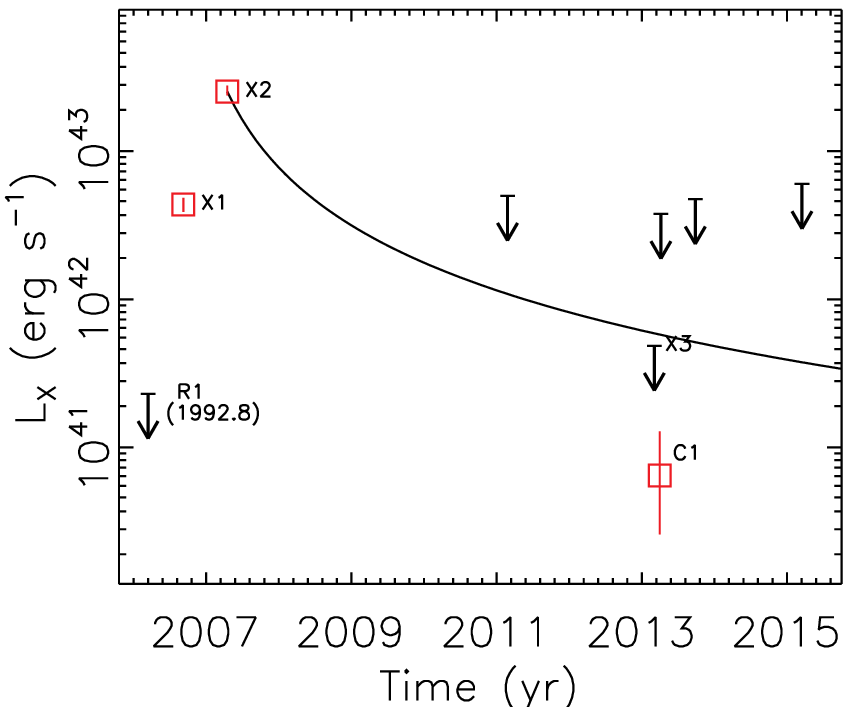}
  }
  \subfigure{\includegraphics[width=3.4in]{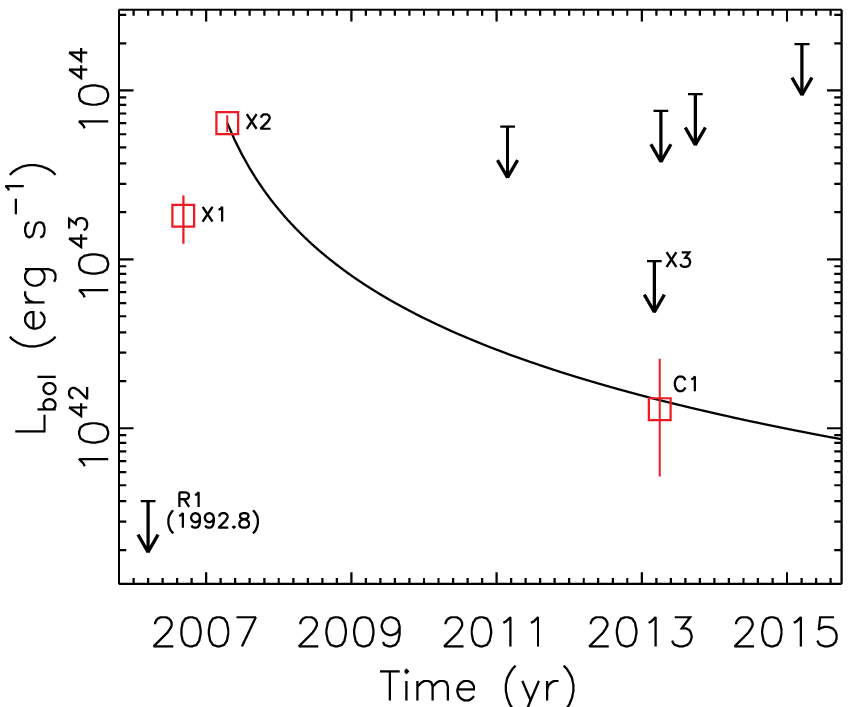}
    }
\caption{The long-term evolution of the unabsorbed X-ray (0.2--10 keV, upper panel) and bolometric (lower panel) luminosities of J1847. The squares with 90\% errors are detections, while the arrows represent $3\sigma$ upper limits (see Table~\ref{tbl:obslog} and Section~\ref{sec:ltxraycurve} for more details). All observations are noted, except those from \emph{Swift}. The solid line in both panel plots a typical TDE evolution curve $(t-t_\mathrm{D})^{-5/3}$, with $t_\mathrm{D}$ assumed to be one month before X1 and with X2 forced to be on the curve.   \label{fig:ltlc}}
\end{figure}

\begin{figure*}
\begin{center}
\includegraphics[width=5.2in]{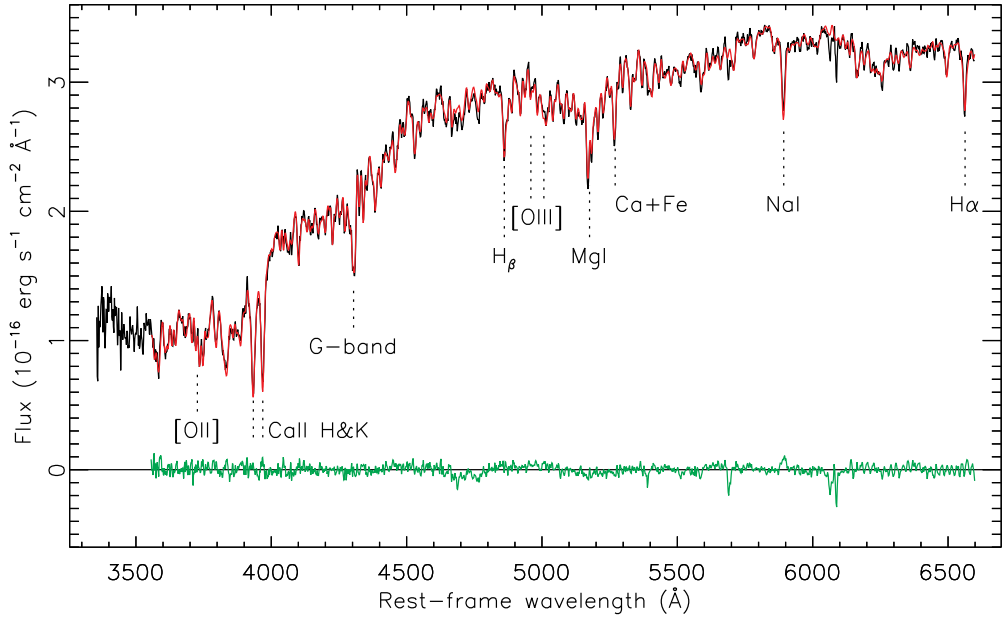}
\end{center}
\vskip -0.2in
\caption{The SOAR optical/UV spectrum of the host galaxy of J1847
  taken in 2013, showing no emission lines but typical stellar
  absorption features.  The \textsc{ppxf} fit is shown as a red solid line, and
  the fit residuals are shown as a green solid line.  \label{fig:soarsp}}
\end{figure*}

\begin{figure*}
\begin{center}
\includegraphics[width=5.2in]{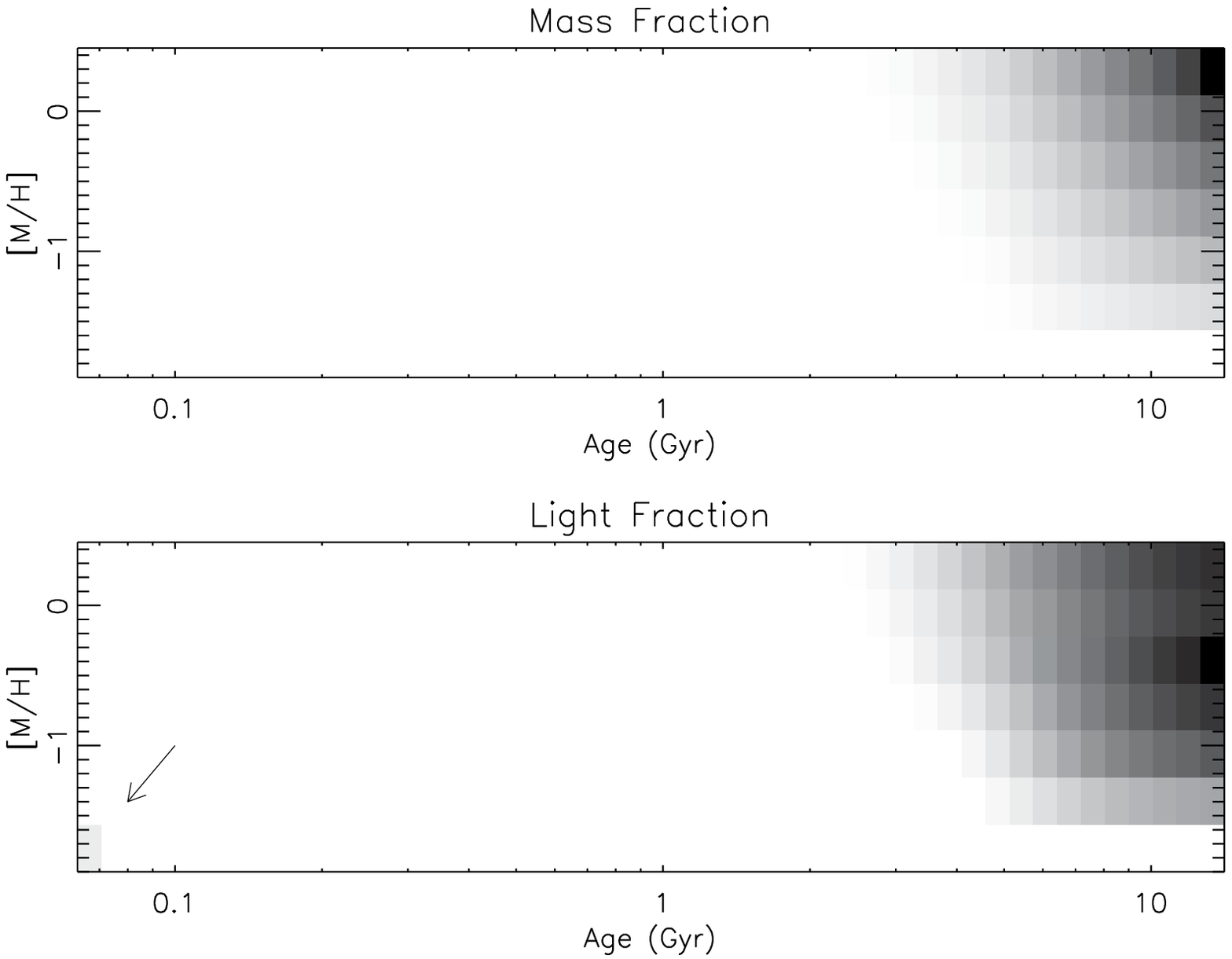}
\end{center}
\vskip -0.2in
\caption{Relative mass and light fractions of different stellar
  populations in the host galaxy of J1847 with respect to
  metallicity and age. Darker shading indicates a larger mass/light
  fraction in the best-fitting model. The light was integrated over 3540--6600 \AA.
The plot shows that the host galaxy is dominated by very old ($>$10 Gyr) populations,  but a very small young ($<$70 Myr) population (pointed with an arrow in the lower panel) seems to be present as well.
  \label{fig:masslightdis}}
\end{figure*}

Figure~\ref{fig:ltlc} shows the long-term evolution of the
  unabsorbed X-ray luminosity $L_\mathrm{X}$ (upper panel) of J1847 in
  0.2--10 keV and the bolometric luminosity $L_\mathrm{bol}$ (lower
  panel). For X1 and X2, we used the luminosities inferred in Lin11
  based on the WABS*SIMPL(DISKBB)) model. Although the possible
  contamination of Source 2 on J1847 was not taken into account when
  we fitted the spectra of J1847 in X1 and X2, we found that the
  spectral fits obtained in Lin11 were hardly affected even if we
  included in the spectral fits an additional weak hard PL component
  with $\Gamma=0.0$ as obtained in our fit to the spectrum of Source 2
  in C1. The upper limit of the luminosity in the \emph{ROSAT}
  observation was estimated assuming a typical spectrum of an active
  galactic nucleus (AGN), i.e., an absorbed PL of
  $N_\mathrm{H}=8.5\times10^{20}$ cm$^{-2}$ (Lin11) and
  $\Gamma_\mathrm{PL}=2.0$ \citep{liweba2012}. For all the other
  observations, we assumed the WABS*SIMPL(DISKBB)) fit to X1 with the
  disk temperature adjusted so that $L_\mathrm{bol}$ of the model
  matches that predicted in the standard TDE model (solid line in the
  lower panel in Figure~\ref{fig:ltlc}), as we did for C1 in
  Section~\ref{sec:xraypos}.  Except for X1, X2, and C1, in which the
source was detected and had little contamination from Source 2, we
calculated the $3\sigma$ upper limits for all other observations
(Table~\ref{tbl:obslog}). For X3, though J1847 was also detected in
this observation, it was most likely seriously contaminated by Source
2. We therefore treated the total detected flux of Source 2 and J1847
as the upper limit of J1847 for this observation. We note that S2 and
S3 are close in time and were thus combined to calculate the upper
limit. This is also the case for S4 and S5.

The source clearly displayed a large variability in $L_\mathrm{X}$, by
a factor of $\sim$430 between X2 and C1, which were separated by
$\sim$6 yr. The \emph{Swift} observations are relatively shallow,
and the non-detection of our source in these observations suggests that
$L_\mathrm{X}$ might have decayed by a factor of $\gtrsim$5 from X2
since 2011. The variability of $L_\mathrm{bol}$ is somewhat less
dramatic (a factor of 49 between X2 and C1) than that of
$L_\mathrm{X}$. This is due a larger bolometric correction for C1 than
for X2. The C1 luminosity was estimated assuming a model of a much
cooler disk ($kT_\mathrm{disk}=35$ eV, see Section~\ref{sec:xraypos})
and thus a smaller fraction of photons in X-rays than the model used
for X2 ($kT_\mathrm{disk}=93$ eV, Lin11).

\subsection{Long-term Optical and UV Evolution}
\label{sec:optuv}

Figure~\ref{fig:soarsp} shows the SOAR spectrum of the host galaxy of
J1847. It exhibits no emission lines but typical stellar absorption
features, as observed in the Gemini spectrum in 2011. The SOAR
spectrum has much higher S/N than the Gemini spectrum and can thus
provide a much better constraint on the level of the nuclear
activity. We obtained the 3$\sigma$ upper limit of the luminosity of
the [O\,\textsc{iii}] $\lambda$5007 emission line to be
$4.9\times10^{37}$ erg s$^{-1}$. This limit has been corrected for
Galactic reddening $\mathrm{E}(B-V)_\mathrm{G}=0.10$ mag
\citep{scfida1998}, but not the intrinsic reddening, which is probably
very small given that the inferred column density from the X-ray
spectral fit is close to the Galactic value (Lin11). Using the
[O\,\textsc{iii}] $\lambda$5007 and 2--10 keV luminosity relation in
\citet{labima2009}, whose dispersion is 0.63 dex, we estimated the
3$\sigma$ upper limit of the persistent unabsorbed 2--10 keV
luminosity to be $4.2\times10^{39}$~erg~s$^{-1}$, or
$1.0\times10^{40}$~erg~s$^{-1}$ in 0.2--10 keV, assuming a PL of
$\Gamma_\mathrm{PL}=2.0$. This upper limit is a factor of $\sim$2700
lower than the peak luminosity (in X2).

Our \textsc{ppxf} fit of the spectrum over 3540--6600 \AA\ (the source
rest frame) is shown in Figure~\ref{fig:soarsp}. The fit inferred the
intrinsic stellar velocity dispersion to be $118\pm14$~km~s$^{-1}$,
which should be taken with caution because it is lower than the
instrumental resolution of $\sim$170~km~s$^{-1}$. The total stellar
mass is $\sim$$2.3\times10^{10}$~\Msun, and the total luminosity
within the fitting band is $\sim$$1.7\times10^{9}$~\Lsun, after
correcting for the slit loss by matching the spectrum to the
integrated $V$-band flux of the host galaxy obtained in
Lin11. Therefore, the \textsc{ppxf} fit suggests a relatively low-mass
galaxy.

The mass and light distributions of the stellar populations with
respect to the age and the metallicity are shown in
Figure~\ref{fig:masslightdis}.  The light-weighted age is 9.6 Gyr, and
the mass-weighted age is 10.3 Gyr, suggesting a galaxy dominated by
old stellar populations.

The UV photometry of J1847 is given in Table~\ref{tbl:obsloguv}. Our
source fell in the FOV of the Optical Monitor
\citep[OM,][]{mabrmu2001} in X1 and X3, but not in X2. Both X1 and X3
have been included in the third release of the XMM-OM Serendipitous
Ultraviolet Source Survey Catalog \citep[SUSS3,][]{pabrta2012}, which
compiles all optical and UV sources detected in the OM between 2000
March and 2015 July, and the magnitudes of our source in
Table~\ref{tbl:obsloguv} are from this catalog. Both
observations suggest the presence of a UV source at the position of
J1847 (Figure~\ref{fig:optimg}). This UV source appeared blue and
variable. Both the $UVW1$ and $UVM2$ fluxes decreased from X1 to X3 by
0.6 mag at the $5\sigma$ and $4\sigma$ confidence levels,
respectively. Because X1 was near the peak of the X-ray flare and had
a much higher X-ray flux than X3, the UV enhancement in X1 is probably
associated with the X-ray flare. The UV photometry from the
\emph{Swift} observations (Table~\ref{tbl:obsloguv}) supports the
presence of a blue UV source at the position of J1847, but it showed
no clear variation in \emph{Swift} observations. We note that the UV
filter set in the UVOT is different from that of the \emph{XMM-Newton}
OM, so their photometry should not be directly compared with each
other.

In the SUSS3 catalog, the UV source was indicated as an extended
source in $UVW1$ and $UVM2$ in X3 but not in X1. To check whether the
UV source was extended in the \emph{Swift} observations, we run the
\texttt{uvotdetect} task in the \textsc{ftools} in the stacked images, one for
each filter ($UVW1$ and $UVW2$). The task gave the profile root mean
squares along the major and minor axises of the sources, from which we
calculated a geometric mean $r$. From $UVW1$, we obtained
$r=1.07\pm0.03$~arcsec for point sources with similar magnitudes as
J1847 and $r=1.41$~arcsec for J1847, clearly suggesting that the UV
source was extended (at the $10\sigma$ confidence level). For $UVW2$,
we obtained $r=1.13\pm0.03$~arcsec for point sources and
$r=1.23$~arcsec for J1847. Therefore in this filter, there is some
sign that the UV source was also extended, at the $3\sigma$ confidence
level. In summary, both the X3 and \emph{Swift} images suggest the
extended nature of the UV source at the late time after the X-ray
emission became very weak. We note that the UV source appeared
symmetric and that there seemed to be no UV emission at the position
of Source 2. Therefore the extended nature of the UV source should not
be caused by any contamination from Source 2.

\section{DISCUSSION AND CONCLUSIONS}
\label{sec:discussion}
Our multiwavelength follow-up observations of J1847 support the TDE
explanation of the source in several aspects. First, the X-ray
monitorings of various observatories confirm the large amplitude of
the X-ray outburst in 2006--2007, with the absorbed X-ray flux decaying by a
factor of $\sim$1000 six years after X2. Secondly, the optical
spectrum of the host galaxy from the SOAR poses a very tight
constraint on the persistent nuclear activity (persistent X-ray
luminosity inferred to be $\lesssim10^{40}$~erg~s$^{-1}$), as
supported by the X-ray observation C1. Thirdly, the C1 observation
supports the nuclear origin of the event ($<0.6$ kpc).

Another support for the TDE explanation for J1847 is that the
  X-ray flux in C1 is fully consistent with that expected for a
  standard TDE, after properly taking into account the shifting of
  photon energies out of the X-ray band due to the cooling of the disk
  (lower panel in Figure~\ref{fig:ltlc}, Sections~\ref{sec:xraypos}
  and \ref{sec:ltxraycurve}). The main result of this band-limited
  effect is that the X-ray luminosity has a steeper decay than the
  standard TDE trend of $t^{-5/3}$, as shown in the upper panel in
  Figure~\ref{fig:ltlc}. \citet{loro2011} showed that the X-ray
  spectral density at 0.2 keV could have an exponential decay, thus
  much steeper than the $t^{-5/3}$ decay, around a year after the
  stellar disruption. We note that our speculation for the presence of
  a strong cool disk in C1 and other follow-up observations does not
  conflict with our lack of the detection of the UV variability in
  \emph{Swift} observations, because the UV emission from such a cool
  disk is expected to be an order of magnitude fainter than the
  extended UV emission detected, which could be due to a small young
  stellar population (see below).

There are other cases in which X-ray flux could have a steeper decay than
$t^{-5/3}$. One is partial disruption of a centrally concentrated star. In this case, a large drop in the decay of the mass accretion rate could occur
  in the months after the flare peak \citep{gura2013}. Temporary large
  drops in the light curves of TDEs could also be caused by the
  presence of a secondary SMBH \citep{lilich2009}, as is probably the
  case for the TDE SDSS J120136.02+300305.5
  \citep{liliko2014}. We cannot completely rule out these
  explanations for the steeper decay of the X-ray flux than $t^{-5/3}$
  in J1847 due to the poor sampling of the light curve.

We have mainly considered the faint emission of J1847 in C1 as due to
the TDE, but we cannot rule it out as the pre-existing very faint AGN
instead. The 0.2--10 keV luminosity of J1847 in C1
($\sim$$1.2\times10^{40}$~erg~s$^{-1}$, assuming an absorbed PL of
$N_\mathrm{H}=8.5\times10^{20}$ cm$^{-2}$ and
$\Gamma_\mathrm{PL}=2.0$, typical of AGN spectra) is compatible with
the $3\sigma$ upper limit ($1.0\times10^{40}$~erg~s$^{-1}$) set by the
SOAR optical spectrum. The emission in C1 is less likely to be due to the
  star-forming activity. The star-formation rate of the host galaxy is
  about 0.2 \Msun\ yr$^{-1}$ based on the $UVW2$ flux \citep[corrected
  for the Galactic reddening,][]{hibuin2003}. This rate implies a
  0.5--10 keV luminosity of $10^{39}$~erg~s$^{-1}$ \citep{racose2003},
  significantly lower than observed in C1.

There is a blue UV source at the nucleus of the host galaxy of
J1847. Given its probably extended
nature in the observations after 2011 (X3 and S1--S6), this UV source
could be due to the presence of a small young stellar population, as
inferred in the \textsc{ppxf} fit to the SOAR spectrum (see the lower panel in
Figure~\ref{fig:masslightdis}). The probable decay of the UV source
from X1 to X3 by $\sim0.6$ mag, however, requires significant
contribution to the UV emission from the TDE in X1. The presence of a
young stellar population at the nucleus is interesting, as it has been
suspected that the TDE rate might be enhanced in young stellar
environments \citep{argasu2014,taspro2017}. Given the dominant very
old stellar populations of the host, one possible explanation for the
nuclear young stellar population might be a recent minor merger. The
possible evidence of this minor merger could be the prominence to the
north of J1847 in the VLT image (Figure~\ref{fig:optimg}).

In Lin11, we ruled out J1847 as AGNs (based on the very soft X-ray
spectra of J1847), foreground stars \citep[based on the large
  X-ray-to-IR flux ratio of J1847, see also][]{liweba2012} and
Galactic supersoft X-ray sources (J1847 would be too faint to be such
an object if it is in our Galaxy). The very large X-ray variability
inferred from our follow-up observations and the extremely weak
nuclear activity as indicated by the SOAR optical spectrum strengthen
our arguments against the AGN explanation for J1847. J1847 is unlikely
to be a $\gamma$-ray burst (GRB) or a supernova (SN), based on typical
arguments applied to most X-ray TDEs of very soft X-ray spectra and
high luminosities over a long time
\citep[e.g.,][]{maliir2014,limair2015,licawe2016}. GRBs and SNe
typically have very hard X-ray spectra, with
$\Gamma_\mathrm{PL}\lesssim$2
\citep{im2007,lereme2013,grnove2013}. SNe could show luminous very
soft X-ray spectra during the prompt shock breakouts
\citep{sobepa2008}, but this phase is expected to be short in time
\citep[less than hours,][]{nasa2012}, while J1847 at least stayed
bright and super-soft for 0.6 yr. Some ultralong GRBs exhibited
relatively soft late-time X-ray spectra
($\Gamma_\mathrm{PL}\lesssim5$), but they normally also had
significant very hard X-ray emission above 2 keV \citep[e.g.,
][]{pitrge2014,magula2015}, but J1847 showed little emission above 2
keV and was much softer in X1 and X2, with $\Gamma_\mathrm{PL}=5.9$
and 6.9, respectively, when fitted with an absorbed PL.

Our \emph{Chandra} observation C1 revealed a faint source (Source 2)
only 6~arcsec away from J1847. This source appears to be in the
galactic plane of the host galaxy of J1847, but no clear optical
counterpart is seen in the VLT image (Figure~\ref{fig:optimg}). If it
is in this galaxy, it would have an absorbed 0.3--8 keV luminosity of
$7\times10^{40}$~erg~s$^{-1}$, fulfilling the definition of an
ultraluminous X-ray source
\citep[$>10^{39}$~erg~s$^{-1}$,][]{feso2011}. However, it is very rare
to have such a bright off-nuclear X-ray source in an early-type galaxy
\citep{fa2006}. One possible exception could be that there was a minor
merger, with Source 2 corresponding to the tidal stripped nucleus of a
merging dwarf galaxy \citep[see][]{fasepf2012,licawe2016}. Although
this explanation seems interesting, given that a minor merger could
also explain the presence of a nuclear young stellar population (see
above), it is still fairly speculative without high-resolution imaging
of the environment.  An alternative explanation for Source 2 is a
background AGN. No matter whether it is an offset nucleus or a
background AGN, it might have a Seyfert 2 type, given its very hard
X-ray spectrum \citep{liweba2012}. We do not favor the explanation of
Source 2 as a $\gamma$-ray burst or a supernova, whose X-ray spectra
typically have $\Gamma_\mathrm{PL}\gtrsim1.0$ \citep{grnove2013}, not as hard as
Source 2. We note that we cannot rule out Source 2 as a foreground
Galactic source either. With a very hard X-ray spectrum, it could be
an intermediate polar \citep{muarba2004}.

\section*{Acknowledgments}

DL is supported by the National Aeronautics and Space Administration
through Chandra Award Number DD3-14066A issued by the Chandra X-ray
Observatory Center, which is operated by the Smithsonian Astrophysical
Observatory for and on behalf of the National Aeronautics Space
Administration under contract NAS8-03060, and by the National
Aeronautics and Space Administration ADAP grant NNX17AJ57G. JS acknowledges
support from the Packard Foundation. We want to thank the former
\textit{Swift} PI Neil Gehrels for approving our ToO request to make
several observations of 2XMMi~J184725.1-631724. This paper is
partially based on observations obtained at the Southern Astrophysical
Research (SOAR) telescope, which is a joint project of the
Minist\'{e}rio da Ci\^{e}ncia, Tecnologia, e Inova\c{c}\~{a}o (MCTI)
da Rep\'{u}blica Federativa do Brasil, the U.S. National Optical
Astronomy Observatory (NOAO), the University of North Carolina at
Chapel Hill (UNC), and Michigan State University (MSU).

\section*{REFERENCES}
\bibliographystyle{mn2e}

\bsp

\label{lastpage}

\end{document}